\documentclass{article}

\usepackage{microtype}
\usepackage{graphicx}
\usepackage{multirow}
\usepackage{booktabs} 
\usepackage{subcaption}
\usepackage{hyperref}

\usepackage{enumitem}
\usepackage[accepted]{mlsys2025}
\usepackage{algorithmic}

\mlsystitlerunning{TruncFormer}

\begin{document}

\twocolumn[
\mlsystitle{TruncFormer: \\
Private LLM Inference Using Only Truncations}



\mlsyssetsymbol{equal}{*}

\begin{mlsysauthorlist}
\mlsysauthor{Patrick Yubeaton}{nyu}
\mlsysauthor{Jianqiao Cambridge Mo}{nyu}
\mlsysauthor{Karthik Garimella}{nyu}
\mlsysauthor{Nandan Kumar Jha}{nyu}
\mlsysauthor{Brandon Reagen}{nyu}
\mlsysauthor{Chinmay Hegde}{nyu}
\mlsysauthor{Siddharth Garg}{nyu}
\end{mlsysauthorlist}

\mlsysaffiliation{nyu}{Department of Electrical and Comptuer Engineering, New York University}
\mlsyscorrespondingauthor{Patrick Yubeaton}{wpy2004@nyu.edu}

\mlsyskeywords{Machine Learning, MLSys}

\vskip 0.3in

\begin{abstract}
  Private inference (PI) serves an important role in guaranteeing the privacy of user data when interfacing with proprietary machine learning models such as LLMs. However, PI remains practically intractable due to the massive latency costs associated with nonlinear functions present in LLMs. Existing works have focused on improving latency of specific LLM nonlinearities (such as the Softmax, or the GeLU) via approximations. However, new types of nonlinearities are regularly introduced with new LLM architectures, and this has led to a constant game of catch-up where PI researchers attempt to optimize the newest nonlinear function. We introduce TruncFormer, a framework for taking any LLM and transforming it into a plaintext emulation of PI. Our framework leverages the fact that nonlinearities in LLMs are differentiable and can be accurately approximated with a sequence of additions, multiplications, and truncations. Further, we decouple the add/multiply and truncation operations, and statically determine where truncations should be inserted based on a given field size and input representation size. This leads to latency improvements over existing cryptographic protocols that enforce truncation after every multiplication operation. We open source our code for community use.
\end{abstract}
]



\printAffiliationsAndNotice{}  
\section{Introduction}
Regular use of transformer-based large language models (LLMs) such as ChatGPT has skyrocketed in recent years. Many of these models are proprietary and accessible via API calls where the user is required to send their input data (such as the prompt, or other context) to the model provider to get inference results. Therefore, these models are not usable if the query involves confidential data. One way to overcome this concern is the use of private inference (PI) with secure multiparty computation (MPC). PI is a process which uses cryptographic techniques to perform machine learning inference where neither the data holder, nor the model owner learn about each other's data. 

Although the possibility of PI with MPC has been demonstrated in numerous previous works, the practicality of PI is still up to debate due to high latency costs~\cite{evans2018pragmatic,liu2017oblivious}. 
At a high level, the latency cost of PI can be broken up into two parts: \emph{in-field} operations and \emph{out-of-field} operations. Traditional cryptographic operations are performed in a field of a fixed size, where the only operations supported by this field are addition and multiplication. These operations are slower than their plaintext variants, but do not serve as the main contributor to latency in PI~\cite{ghodsi2020cryptonas}. Out-of-field operations such as division, exponentiation, ReLU, and many others are the major source of additional latency in PI, and therefore most existing PI protocols used in modern AI models must use  techniques such as Yao's garbled circuit~\cite{yao1986generate} or Oblivious Transfer~\cite{kilian1988founding}. However, both of these techniques are communication-heavy and thus add significant latency to practical PI implementations. 

To counter the large latency incurred by out-of-field PI operations, a large body of work has emerged. Earlier works such as Delphi~\cite{mishra2020delphi} and Circa~\cite{ghodsi2021circa} create polynomial approximations for out-of-field operations. Later approaches such as CryptoNAS~\cite{ghodsi2020cryptonas} and   Sphynx~\cite{cho2022sphynx}, SNL~\cite{cho2022selective}, and SE-Net~\cite{kundu2023learning} adapt network architectures with a minimal number of out-of-field operations. However, most of these methods have considered computer vision applications involving convolutional/residual networks, and focus on the ReLU as the sole out-of-field operation. 

\begin{table*}[!t]
    \scriptsize
    \centering
    \resizebox{0.8\textwidth}{!}{
    \begin{tabular}{ccc}\toprule
    Approach & Method & Operation Optimized \\ \midrule
    Iron~\cite{hao2022iron} & New PI protocol & GeLU, Softmax, MatMul \\
    MPCFormer~\cite{li2022mpcformer} & Approximation,  KD & GeLU, Softmax \\
    BumbleBee~\cite{lu2023bumblebee} & Approximation, new protocol & GeLU, Softmax, MatMul \\
    CipherGPT~\cite{hou2023ciphergpt} & New protocol & GeLU, MatMul, top-k sampling\\
    \textsc{Puma}~\cite{dong2023puma} & Approximation, new protocol & GeLU, Softmax, Layer Norm\\
    TruncFormer (our method) & Static truncations & Truncation\\ \bottomrule
    \end{tabular}
    }
    \caption{\sl Overview of the major works in LLM private inference (PI). Most existing methods reduce PI latency by either adapting the model architecture with numerical approximations of nonlinearities, or improving the crypto protocol. Our method, TruncFormer, emulates all LLM operations using a sequence of adds, multiplies, and (statically chosen) truncations.}
    \label{tab:rel_work}
\end{table*}

With the rise of LLMs and the transformer architecture, PI methods now have to contend with new families of out-of-field operations. This has led to a small, but growing body of work that focuses on optimizing specific operations such as the softmax or the GeLU. More recently, MPCFormer~\cite{li2022mpcformer},  PriViT~\cite{dhyani2023privit}, MPCViT~\cite{zeng2022mpcvit}, and Sal-ViT~\cite{zhang2023sal} have employed similar network adaptation ideas to reduce the burden of cryptographically processing nonlinear operations within transformer models. Other related works in this direction include~\cite{hao2022iron,hou2023ciphergpt,zheng2023primer}; see~\cite{luo2024secformer} for a comprehensive survey. However, because of the rapid progress in LLM research, there is a constant race to devise new protocols, or new PI-friendly approximations of nonlinear operations. 

\paragraph{Our contributions.} In this paper, we introduce \emph{TruncFormer}, a new framework for LLM inference that can be used \emph{in conjunction} with any network architecture and any PI protocol.  Specifically: 

\begin{enumerate}[leftmargin=*,nosep]
    \item TruncFormer takes \emph{any} transformer  model and for a given field size and a bit-width for arithmetic operations, and statically determines where to compute truncations wherever overflows can occur. 
    \item TruncFormer modifies various Crypten~\cite{knott2021crypten} approximations for out-of-field operations so that they are suitable for use in today's public LLMs.
    \item Putting these together, we show that TruncFormer accomplishes LLM private inference with in-field operations (adds/multiplies) and truncations alone.
    \item We validate TruncFormer using a plaintext-library that accurately emulates the costs (latency and accuracy) of implementing our approach in the PI setting for two popular LLMs (Llama-7B, Gemma-2B). We open-source this library for community use (see \url{https://anonymous.4open.science/r/fixed_llm-D334/README.md}).
\end{enumerate}

\paragraph{Techniques.} TruncFormer rests upon two straightforward but key insights. Our first key insight is that nonlinearities in LLMs (softmaxes; layer norms; activations such as the GeLU, SiLU, SwiGLU, etc) are \emph{differentiable}, and can be accurately represented with iterative approximations such as Newton-Raphson or Taylor series. Therefore, all computation within an LLM can be explicitly written in terms of (a long chain of) in-field operations. 

However, this comes at a price, since ciphertext operations are performed over a field of a fixed size, and one has to be conscious of the limits of this size parameter. Multiplication between an input represented with $m$-bits and an input represented with $n$-bits can lead to an output with $(m+n)$-bits of size. Therefore, one must ensure that $(m+n)$ bits is less than or equal to the number of bits used to define the maximum number in our field; if not, one must \emph{truncate} the output by dropping least significant bits. \emph{Crucially, such a truncation is an out-of-field operation} as pointed out in~\cite{li2023efficient}; this fact seems to have been (relatively) unaddressed by most previous works in this area. 

\begin{figure}[h]
    \centering
    \begin{subfigure}{\columnwidth}
        \centering
        \includegraphics[width=\columnwidth]{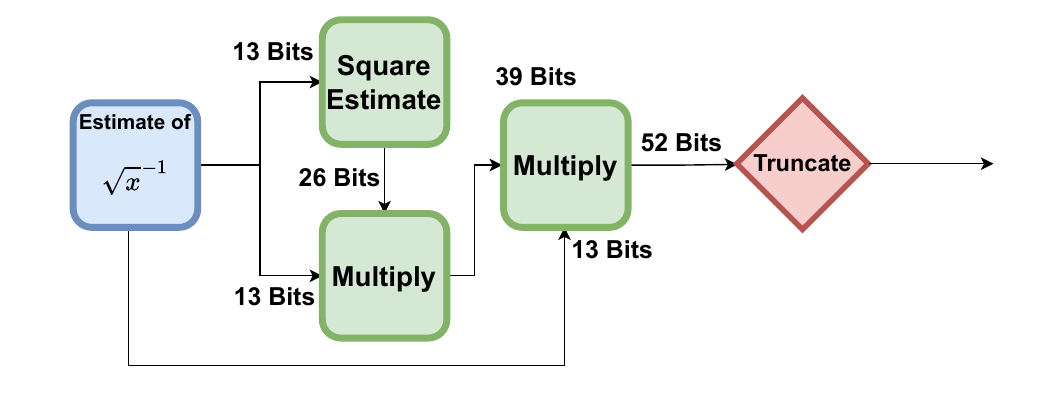} 
        \caption{TruncFormer with a 64 bit field.}
        \label{fig:figure1}
    \end{subfigure}
    
    \vspace{0.1in} 

    \begin{subfigure}{\columnwidth}
        \centering
        \includegraphics[width=\columnwidth]{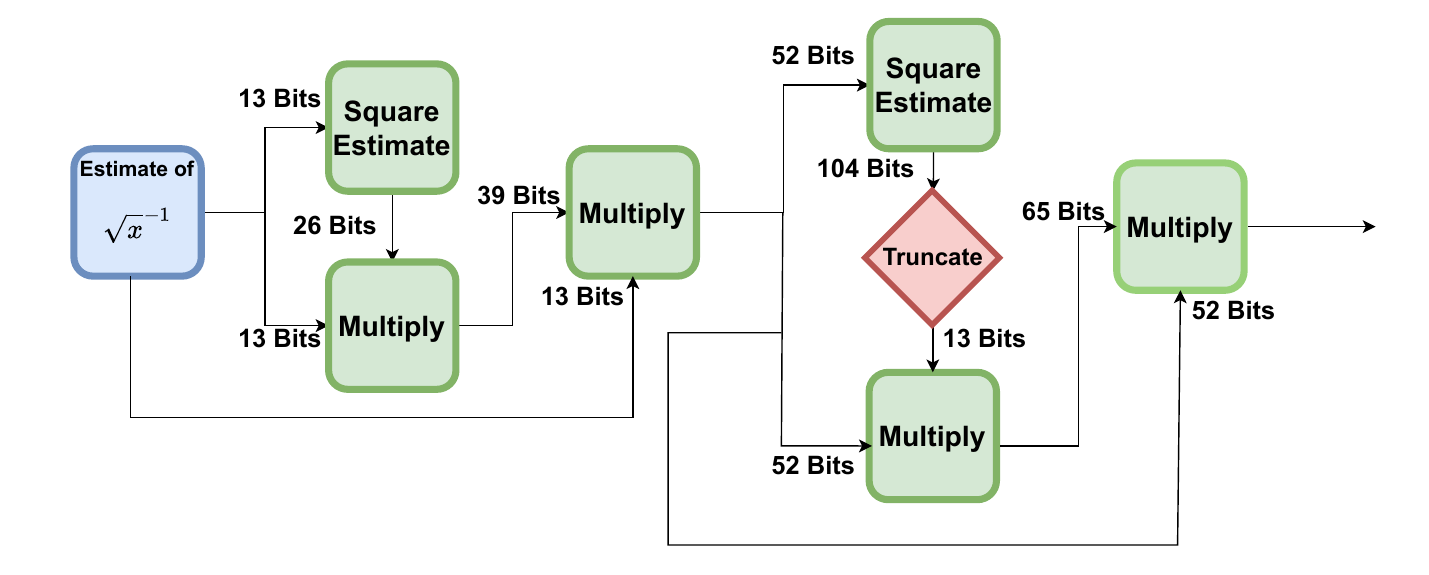}
        \caption{TruncFormer with a 128 bit field.}
        \label{fig:figure2}
    \end{subfigure}

    \begin{subfigure}{\columnwidth}
        \centering
        \includegraphics[width=\columnwidth]{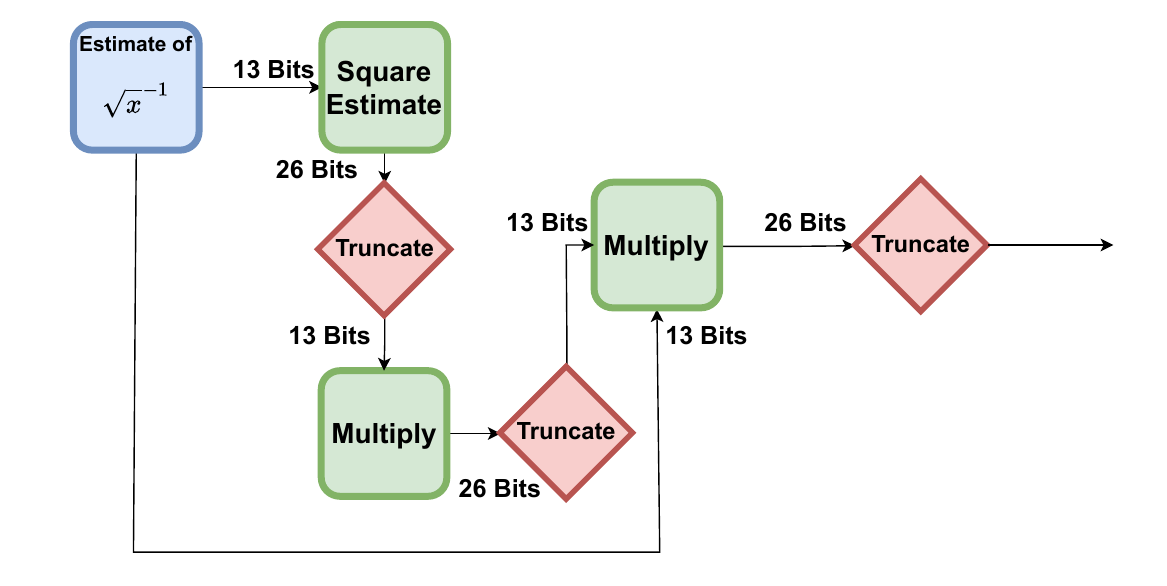} 
        \caption{Existing private inference protocol such as CrypTen.}
        \label{fig:figure3}
    \end{subfigure}
    
    \caption{Newton Raphson flowchart for the approximation of the inverse squareroot. Standard private inference protocols make use of a truncation after every operation which we show is excessive.}
    \label{fig:tldr_figure}
\end{figure}

To our knowledge, all public implementations of PI methods combine with every in-field operation (multiplication) with one out-of-field operation (truncation) to stay within the fixed field size. Our second key insight is that this is unnecessary: for a given fixed point  representation and field size, TruncFormer examines the LLM architecture to statically determine which in-field operations might overflow (and therefore need truncations), and only applies truncations after such operations. In this manner, we show that tremendous savings can be obtained if we judiciously apply truncations only when necessary. This high level idea for TruncFormer can be seen in Figure~\ref{fig:tldr_figure}. We report extensive experiments on modern public LLMs (Llama-7B, Gemma-2B) and demonstrate significant improvements over the  state-of-the-art for LLM inference, \textsc{Puma}~\cite{dong2023puma}. 

\section{Related Work}
\paragraph{Cryptographic Primitives}
The major cryptographic techniques used in PI are secret sharing~\cite{brickell1989some}, homomorphic encryption~\cite{yi2014homomorphic}, garbled circuits~\cite{yao1986generate}, and oblivious transfer~\cite{liu2017oblivious}. A combination of secret sharing and homomorphic encryption is used to compute in-field operations. Garbled circuites~\cite{yao1986generate} and oblivious transfer~\cite{liu2017oblivious} are the major methods of performing out-of-field operations.

We will follow the Delphi protocol~\cite{mishra2020delphi}. This protocol assumes that both parties are honest-but-curious which means that both parties will follow the protocol, but will attempt to learn information about the other party's input. Delphi uses secret sharing and homomorphic encryption for in-field operations. They use garbled circuits for out-of-field operations. The method we propose below will be applicable to other major protocols,  such as the SEMI2K protocol found in SecretFlow~\cite{ma2023secretflow}.

\paragraph{Private Inference}
PI works before the popularization of transformers  were largely focused around ConvNets~\cite{mishra2020delphi} and ResNets~\cite{ghodsi2020cryptonas}, with ReLUs being the only nonlinearities to cause latency bottlenecks. However, PI works after the advent of transformers have had to deal with many more out-of-field operations. These include the softmax, GeLU, SiLU, inverse square-root, among many others. In addition, transformers are filled with very large matrix multiplications, further adding to PI latency. 



With the advent of the transformer, new techniques have emerged. 
Iron~\cite{hao2022iron} designs MPC protocols specifically for transformer-based architectures. They design secure protocols for matrix multiplication, softmax, GeLU, and layer norm. MPCFormer~\cite{li2022mpcformer} introduce the idea of aggressive approximations for out-of-field operations. They then regain model accuracy with knowledge distillation. CipherGPT~\cite{hou2023ciphergpt} focus on optimizing the matrix multiplication protocol to improve end to end latency. BumbleBee~\cite{lu2023bumblebee} and \textsc{Puma}~\cite{dong2023puma} introduce new approximations for the GeLU and softmax that retain accuracy while also avoiding the fine-tuning phase from MPCFormer. A summary of these works can be found in Table~\ref{tab:rel_work}.

The vast majority of works in this field centered around LLMs have focused on making incremental improvements through improved and optimized approximations. However, these optimizations have been focused on the most common activation functions such as GeLU. New activation functions such as GLU and SwiGLU~\cite{shazeer2020glu} are being utilized in new transformer architectures such as Llama 2~\cite{touvron2023llama2}. Therefore, PI researchers will be in a constant race to optimize the newest activation function. Despite this focus on optimizing activation functions such as the GeLU and Softmax, PI is still not practical from a latency perspective. Thus, we argue that a new direction is necessary to push PI from a research idea into a practical idea. We show in this paper that truncations are responsible for the vast majority of online latency in PI. In addition, we show that differentiable functions can be represented with only truncations and comparisons as the out-of-field operations. 

\section{The TruncFormer Framework}

\paragraph{Fixed point representations.}  
LLM implementations traditionally represent weights and hidden states as being stored in floating point containers. However, most cryptographic protocols operate under a fixed field which is not easily compatible with floating point representations. Therefore, we first transform the weights and hidden states into an integer fixed-point representation to accurately emulate the setting for cryptographic operations. Our fixed point encoding is performed by multiplying the floating point input by a large scaling factor $M$ and then rounding to the nearest integer. We determine $M$ by choosing a desired precision in terms of bits such that $M=2^n$ where $n$ is the desired number of bits of precision. To convert this fixed point number back to a floating point number, we simply divide the fixed point number by $M$.

\paragraph{Statically determined truncations.}
Exceeding field size in a fixed-point representation will lead to overflow errors, which will degrade the performance of the machine learning model. Therefore, truncations are necessary to keep our input representations smaller than the field size. To our knowledge, all publicly available implementations of PI solutions for machine learning perform a truncation after every operation that can change bit-width, such as multiplication. We posit that this is overly conservative and leads to high latency costs. 

Therefore, our method proceeds as follows. Given a directed acyclic graph (DAG) of addition and multiplication operations, one can statically determine the optimal locations for truncations while maintaining model accuracy. Each vertex in the DAG represents a binary arithmetic operation, with two input edges representing the bit-size of the inputs for that operation. The vertex will have one output edge representing the bit-size of the output for that operation. The method calculates the worst case number of output bits for every bit-altering operation and only insert a truncation if that output would exceed our field size. This process is done \emph{independently of the input} and can therefore be done in a static manner. We detail this in Algorithm~\ref{alg:stat_trunc}.
\begin{algorithm}
\caption{Statically Determining Trunctions}\label{alg:stat_trunc}
\begin{algorithmic}
\STATE Input: $G \leftarrow (V, E)$, $k \leftarrow$ Input size in bits, $\mathcal{F} \leftarrow$ Field size in bits. 
\STATE Output: $L \leftarrow$ List that contains the truncation locations, $H\leftarrow(V', E')$
\WHILE{G is not empty} 
    \STATE Remove first vertex $V\leftarrow (operation, E_1, E_2, E_3)$
    \IF{operation is addition}
        \STATE worst case $\leftarrow max(E_1, E_2) + 1$
    \ELSIF{operation is multiplication} 
        \STATE worst case $\leftarrow E_1 + E_2$
    \ENDIF{}
    \WHILE{worst case $> \mathcal{F}$}
        \IF{$E_1 > E_2$}
            \STATE $E_1 \leftarrow k$
            \STATE Add $E_1$ to L
        \ELSE{}
            \STATE $E_2 \leftarrow k$
            \STATE Add $E_2$ to L
        \ENDIF{}
        \STATE worst case $\leftarrow \max(E_1, E_2) + 1$
    \ENDWHILE{}
    \STATE $E_3 \leftarrow$ worst case
    \STATE $H \leftarrow (V,H)$.
\ENDWHILE
\STATE \textbf{return} L, H
\end{algorithmic}
\end{algorithm}

\paragraph{Approximations for common out-of-field operations}
 Nonlinear operations such as the softmax, GeLU, SiLU, and others would require very complex garbled circuits (GC) or oblivious transfer (OT) protocols, leading to extremely high latency costs. Therefore, many recent works have focused on creating approximations for these operations that reduce the complexity of the GC/OT protocols needed to perform these operations. Works such as Crypten~\cite{knott2021crypten} have gone even further by reducing these operations down to in-field operations, truncations, and comparisons. This method goes hand in hand with our truncation optimization leading to further optimizations of these out-of-field operations. 

The approximations created by  Crypten~\cite{knott2021crypten} are only stable at a bit precision of 16-bits and thus have not seen wide use in modern PI works. Therefore, we modify parts of these approximations to ensure that they produce accurate results with field sizes up to 128-bits. The majority of these approximations follow either a limit approximation method or the Newton-Raphson method.\footnote{The full details for the CrypTen approximations can be found at this link: \url{https://github.com/facebookresearch/CrypTen/blob/main/crypten/common/functions/approximations.py}.} We showcase the exponential (\ref{eq:exp_approx}) and the inverse square-root (\ref{alg:rsqrt}) algorithms below as examples of how these approximation methods work. 

\begin{equation}\label{eq:exp_approx}
    e^x\approx \left( 1 + \frac{x}{2^8}\right)^{2^8}
\end{equation}

\begin{algorithm}
\caption{Newton-Raphson Approximation of Inverse Square-root}\label{alg:rsqrt}
\begin{algorithmic}
\STATE \textbf{Input}: x, $N_{iter} \leftarrow$ Number of iterations
\STATE \textbf{Output}: Approximation for $\frac{1}{\sqrt{x}}$
\STATE estimate $\leftarrow$ exp$(-(\frac{x}{2}+0.2))\times2.2 + 0.2$
\STATE estimate $\leftarrow$ estimate$\times\frac{1023}{1024}$
\WHILE{iterations  $\le N_{iter}$}
    \STATE estimate $\leftarrow$ estimate$\times(3-x\times\textrm{estimate}^2)/2$
\ENDWHILE{}
\STATE \textbf{return} estimate
\end{algorithmic}
\end{algorithm}

One of the most important aspects of the Newton-Raphson approximations is the initial guess. This not only impacts the number of iterations required for an accurate answer, but also whether an accurate answer can be reached in the first place. We found that the initial guess for Crypten's algorithms often was not low enough to converge to the positive root. Therefore, we lower some of the initial guesses to allow for convergence. However, to ensure that we receive accurate results, we also increase the number of iterations performed. An important question for future work is how best to determine these initial guesses to minimize the number of iterations.

\section{Experiments} \label{sec:experiments}

\subsection{PI Protocol}\label{sec:bench}
For convenience, we follow the Delphi~\cite{mishra2020delphi} protocol for private inference. We use additive secret sharing for in-field operations and garbled circuits (GC) for out-of-field operations. We perform unit tests for matrix multiplications, element-wise multiplications, and truncations to obtain theoretical estimates for PI latency. We found that benchmarking end to end private inference is prohibitively expensive in terms of time and hardware, thus we went for a unit test approach. The latency issue is also shown empirically by Puma where it takes their system over 200 seconds to generate one token from an eight token input on Llama-7B. We detail the unit testing approach below and how it was applied to get the latency results in the later sections.

We benchmark the latency of truncations with the EMP-Toolkit~\cite{emp-toolkit}. We follow the methodology described in~\cite{mo2023haac} for performing the benchmarking. We benchmark a scalar truncation operation for two different field sizes: 64 bits and 128 bits. We truncate the scalar down to a 13 bit representation. For our benchmarking process, we treat the client as the garbler and the server as the evaluator. We recombine the shares, and ensure they are within the field by taking the modulus of the output with respect to the field size. Afterwards, we perform the truncation and then split the output back into secret shares. We benchmark the garbler and evaluator time for field sizes of 64 bits and 128 bits. The results are shown in Table~\ref{tab:gc_trunc}. We use the evaluator times for online latency estimation, since garbling can be performed in a preprocessing phase. 

\begin{table}[ht]
    \scriptsize
    \centering
    \resizebox{0.8\columnwidth}{!}{%
    \begin{tabular}{@{}ccc@{}}\toprule
    Field Size (bits) & Garbler (ms) & Evaluator (ms) \\ \midrule
    64 & 0.098 & 0.099 \\
    128 & 0.152 & 0.175 \\ \bottomrule
    \end{tabular}%
    }
    \caption{\sl Garbled Circuit latency for one scalar truncation at varying field sizes.}    \label{tab:gc_trunc}
\end{table}

We benchmark matrix multiplications and element-wise multiplications by performing unit tests on the beaver triple multiplication formula. These tests are performed on a Lenovo SD650 CPU with 32 GB ram. We unit test multiplications where the matrix sizes match the sizes seen in Llama-7B. We then test various batch sizes for those multiplications to ensure that the time scales linearly. We normalize the multiplication times on a per scalar multiplication basis. For example, a matrix multiplication for matrices of size $[n \times m]$ and $[m \times p]$ would have $n*m*p$ operations. We use these values to estimate the private inference latency caused by matrix multiplications or element-wise multiplications.



However, after performing experiments using the TruncFormer framework, we find that the vast majority of the private inference latency is dedicated to the truncation operation. We see that truncation time takes up anywhere from 90\% to 99\% of the private inference latency. This is highlighted in Table~\ref{tab:mult_trunc}. Therefore, the remainder of the results presented will focus solely on truncation time.

\begin{table}[ht]
\scriptsize
    \resizebox{\columnwidth}{!}{%
        \begin{tabular}{ccccc}\toprule
       HumanEval  &  ARC & GSM8K & MATH & MMLU\\ \midrule
       99.62\%  & 91.13\% & 98.70\% & 98.92\% & 92.22\%\\ \bottomrule
    \end{tabular}%
    }

    \centering
    \caption{\sl Percentage of total PI time dedicated to truncations for various downstream benchmarks on Llama-7B.}
    \label{tab:mult_trunc}
\end{table}

\subsection{Results}

\paragraph{Baseline}
We compare our work to the current state of the art (Puma). However, we note the following for the sake of fair comparisons. The main contributions of \textsc{Puma} are optimizations to the softmax and the GeLU/SiLU. The remaining out-of-field operations they use are sourced directly from the SecretFlow library~\cite{ma2023secretflow}. We cannot directly compare our work on SecretFlow because they do not natively support the decoupling of multiplication and truncation operations. Therefore, we implement the \textsc{Puma} optimizations for softmax and GeLU/SiLU into our framework and use our implementations for the rest of the out-of-field operations.

\begin{table*}[ht]
\scriptsize
\resizebox{\textwidth}{!}{%
        \begin{tabular}{@{}ccc|cc|c|cc@{}} \toprule
        Benchmark & Metric &  Llama-7B & Llama-7B-T & Llama-7B-P & Gemma-2B & Gemma-2B-T & Gemma-2B-P \\ \midrule
        \multirow{2}{*}{Human Eval} & Pass @ 1 & 1.22 & 2.44 & 1.83 & 11.59 & 10.37 & 10.37 \\ \vspace{0.05in}
        & Speed-up ($\times$) & & \textbf{0.99} & &  & \textbf{1.09} & \\        
        \multirow{2}{*}{MBPP} & Pass @ 1 & 17.00 & 18.00 & 18.00 & 33.00 & 32.00 & 31.00 \\
        & Speed-up ($\times$) & & \textbf{1.02} & &  & \textbf{1.02} & \\ \midrule
        \multirow{2}{*}{GSM8K} & Acc (\%) & 10.16 & 9.93 & 9.17 & 17.06 & 15.92 & 16.22 \\ \vspace{0.05in}
        & Speed-up ($\times$) & & \textbf{1.17} & &  & \textbf{1.09} & \\
         \multirow{2}{*}{MATH} & Acc (\%) & 3.54 & 3.54 & 3.29 & 16.34 & 16.09 & 16.09 \\
        & Speed-up ($\times$) & & \textbf{1.13} & &  & \textbf{1.03} & \\  \midrule
         \multirow{2}{*}{ARC-Easy} & Acc (\%) & 72.85 & 73.02 & 72.43 & 72.22 & 72.52 & 71.93 \\\vspace{0.05in}
        & Speed-up ($\times$) & & \textbf{1.09} & &  & \textbf{1.05} & \\ 
         \multirow{2}{*}{HellaSwag} & Acc (\%) & 76.20 & 76.32 & 75.80 & 71.44 & 71.36 & 71.48 \\ \vspace{0.05in}
        & Speed-up ($\times$) & & \textbf{1.16} & &  & \textbf{1.06} & \\ 
         \multirow{2}{*}{MMLU} & Acc (\%) & 29.92 & 29.94 & 30.00 & 32.91 & 32.87 & 32.91 \\ \vspace{0.05in}
        & Speed-up ($\times$) & & \textbf{1.28} & &  & \textbf{1.09} & \\ 
        \multirow{2}{*}{WinoGrande} & Acc (\%) & 69.93 & 70.01 & 69.69 & 64.88 & 64.96 & 65.27 \\ \vspace{0.05in}
        & Speed-up ($\times$) & & \textbf{1.07} & &  & \textbf{1.05} & \\ \midrule
        \multirow{2}{*}{WikiText-2} & Perplexity & 5.04 & 5.03 & 5.06 & 1550.44 & 1886.23 & 1528.87 \\
        & Speed-up ($\times$) & & \textbf{1.92} & &  & \textbf{1.84} & \\         
        \bottomrule
        \end{tabular}%
    }
    \centering
    \caption{\sl Performance of plaintext and private inference models on various downstream benchmarks. Llama-7B and Gemma-2B are the plaintext versions of the models. A suffix of "-T" indicates the TruncFormer version of that model while the "-P" suffix indicates the Puma version of that model. Speed ups are reported by taking the truncation time for the Puma model and dividing that by the truncation time for the TruncFormer model.}
    \label{tab:main_res}
\end{table*}

\begin{table*}[t]
    \centering
      \begin{tabular}{ccc} \toprule
        Benchmark & Evaluation Framework & Settings \\ \midrule
        HumanEval & HumanEval Github~\cite{chen2021codex} & max new tokens=1024\\
        MBPP & Big Code Bench~\cite{zhuo2024bigcodebench}& limit=100, n\_samples=1 \\
        GSM8K & lm evaluation harness~\cite{eval-harness} & tasks=gsm8k\_cot\\
        MATH &lm evaluation harness & tasks=minerva\_math\_algebra\\
        ARC-Easy &lm evaluation harness &tasks=arc\_easy \\
        HellaSwag &lm evaluation harness & tasks=hellaswag\\
        MMLU &lm evaluation harness & tasks=mmlu\\
        WinoGrande &lm evaluation harness & tasks=winogrande\\
        WikiText-2 & Pytorch& Hugging Face Perplexity of Fixed-Length models\\
    \end{tabular}
    \caption{Evaluation frameworks for our benchmarks.}
    \label{tab:eval_framework}
\end{table*}

\paragraph{Downstream Latency Evaluations}
We evaluate our framework by applying it to Llama-7B~\cite{touvron2023llama} and Gemma-2B~\cite{team2024gemma}. We compare the accuracy of the original models, our method, and \textsc{Puma}. The results are present in Table~\ref{tab:main_res}.

We examine coding benchmarks, math benchmarks, common sense reasoning benchmarks, and a perplexity benchmark. HumanEval~\cite{chen2021codex} and MBPP~\cite{austin2021program} are python programming datasets where the LLM must write a python program to answer the question presented in the prompt. We use the entirety (164) of HumanEval and we use a subset (100/500) of MBPP. GSM8K~\cite{cobbe2021training} is a set of grade school math word problems that require multi-step reasoning comprising of 1319 test problems. MATH~\cite{hendrycks2021measuring} is a dataset of math competition problems comprising 12,500 problems from various mathematical domains. We use the minerva math and algebra subset. ARC-Easy~\cite{yadav2019quick}, HellaSwag~\cite{zellers2019hellaswag}, MMLU~\cite{hendrycks2020measuring}, and WinoGrande~\cite{sakaguchi2021winogrande} are common sense reasoning datasets from various domains comprising 9501, 40168, 56168, and 44000 problems respectively. WikiText-2~\cite{merity2016pointer} is a pretraining dataset containing over 100 million tokens. We use the wikitext-2-raw-v1 subset.

We first note that both methods have a comparable accuracy to the original Llama-7B/Gemma-2B on the vast majority of the benchmarks presented. Thus, our approximations have not negatively impacted the model's inference ability. However, the WikiText-2 dataset sticks out because Gemma-2B-T has a perplexity of 1886.23 whereas the original Gemma-2B has a perplexity of 1550.44. We acknowledge that this is an outlier value compared to the other accuracy metrics presented in Table~\ref{tab:main_res} likely caused by an intricacy in Gemma-2B (as seen by its much higher perplexity than Llama-7B). 

We next note that TruncFormer models are consistently on par or faster than their respective Puma models. Some of the benchmarks display very minor speedups such as MBPP whereas others display nearly 2x speedups such as in WikiText-2. To better understand this phenomen, we explore the major sources of truncations in TruncFormer: softmax, GELU/SiLU, Layer Norm, and truncations associated with regular matrix multiplications.


\paragraph{Truncation at the Operation Level}
We provide a breakdown of the latency cost for important nonlinear operations such as the softmax, GELU/SiLU, and Layer Norm. We calculate the percentage of the truncation time that each nonlinear operation's approximation takes for a variety of benchmarks on Llama-7B-T and Gemma-2B-T. The results are shown in Table~\ref{tab:op_break}. 

The WikiText-2 benchmark had the largest speed up of around 1.92x for Llama-7B-T. When looking at the operation breakdown we see that the majority of its inference latency is associated with the softmax operation. MMLU also produced a notable speed up of 1.28x and it has the second highest percentage of time associated with the softmax. In addition to this, we note that both of these benchmarks have similar structures in terms of input size and output generation length. Both of the expected inputs are long, but the output is only expected to generate one token. For MMLU we generate a multiple choice answer whereas for WikiText-2 we generate the next token's logits to calculate the loss. This setting explains the high amount of truncation time spent on the softmax. In a traditional language generation setting we would utilize the KV-Cache to avoid recalculating values computed during previous forward passes. Therefore, the first forward pass is the most intensive for the softmax because it has to perform computations for all input tokens. Therefore, since MMLU and WikiText-2 only generate one token, we have to dedicate a significant amount of time calculating the softmax for all input tokens. 

Although this explains the large softmax time for WikiText-2 and MMLU, it doesn't explain the smaller softmax times for ARC-Easy, HellaSwag, and WinoGrande. These benchmarks are also multiple choice question datasets which output single tokens. Therefore, one might wonder why they have comparatively lower softmax times. For this we have to explore the length of the input tokens. On average the length of a prompt from the datasets will follow this rule: WinoGrande < ARC-Easy < HellaSwag < MMLU < WikiText-2. In addition to this fact, we see that the softmax timings follow the same order. Therefore, it is understandable that these benchmarks would have less softmax time allocated to them despite also having one token outputs. 

Another notable feature of our multiple choice benchmarks is their extremely high amount of GELU/SiLU time. We see that in WinoGrande it goes up to 85\% of the total inference time. This can be explained by the model architecture. In most modern LLMs, the MLP contains the most parameters in the model by a vast majority. One of these parameters also serves as input to the GELU/SiLU for the feed forward network. Due to the increase in parameters, we will see an increase in GELU/SiLU time compared to a nonlinearity with a smaller number of parameters such as the softmax. 

The coding (HumanEval, MBPP) and math (GSM8K, MATH) benchmarks provide us with some interesting questions as well. On average, these benchmarks have a significant amount of their latency dedicated to MatMul truncations rather than softmax or GELU/SiLU. This seems like a confusing result until we look into the actual generations performed by the model. Unlike the single output token benchmarks that were previously discussed, these coding and math benchmarks have token generation lengths ranging from 500 to 2000 tokens. Following this logic we would expect the GELU/SiLU to dominate because an increase in the number of forward passes will always result in the softmax being faster than the GELU/SiLU. However, we see an interesting trend in these generations. The vast majority of these generations are not complete generations that reach 500 - 2000 tokens. In fact, many of them stop after the first token. This behavior (seen in the plaintext and TruncFormer versions of the models) is likely caused by the LLM generating an "end of sequence" token when it isn't able to meaningful respond to a question. Therefore, we will have many generations acting like WikiText-2 where they have a long input sequence and 1 output token. However, there are a few generations where the LLM does generate a significant number of tokens. These generations will contribute to increasing the GELU/SiLU's fraction of the total truncation time. Therefore, when averaging out the times across the whole dataset we see that the softmax and GELU/SiLU have almost "cancelled each other out". The reason why the majority of the time goes to the MatMul truncations is simple: MatMuls will be present in both generation settings and therefore will always have a notable impact on latency.

These observations raise some interesting points about the operation breakdown and its relation to TruncFormer's speed up over Puma. One general trend we see is that as the \% of GELU/SiLU time increases, TruncFormer has closer latency to Puma. This suggests that our approximations for GELU/SiLU may have significantly different latencies. We perform unit tests on these approximations and find that Puma's approximation is about 1.5x faster than TruncFormer's approximation. This explains the close latency values for benchmarks such as ARC-Easy, and WinoGrande. In addition, this shows the strength of TruncFormer's statically determined truncations. Despite the fact that TruncFormer is operating with a significantly slower approximation than Puma, it is able to make up that latency by cleverly choosing truncation locations.


\begin{table}[ht]
\scriptsize
\resizebox{\columnwidth}{!}{%
        \begin{tabular}{@{}cccc@{}} \toprule
        Benchmark & Nonlinearity &  Llama-7B-T & Gemma-2B-T \\ \midrule
        \multirow{4}{*}{Human Eval} & Softmax & 1.96 & 1.00 \\
        & GELU/SiLU & 2.74 & 9.55 \\     
        & Layer Norm & 0.34 & 0.39\\ \vspace{0.05in}
        & MatMul Truncations & 94.97 & 89.09 \\
        \multirow{4}{*}{MBPP} & Softmax & 1.90 & 0.91 \\
        & GELU/SiLU & 1.72 & 7.94 \\     
        & Layer Norm & 0.21 & 0.33\\ \vspace{0.05in}
        & MatMul Truncations & 96.17 & 90.82 \\\midrule
        \multirow{4}{*}{GSM8K} & Softmax & 9.78 & 3.94 \\
        & GELU/SiLU & 9.50 & 24.68 \\     
        & Layer Norm & 1.17 & 1.01\\\vspace{0.05in}
        & MatMul Truncations & 79.55 & 70.36 \\
        \multirow{4}{*}{MATH} & Softmax & 7.43 & 2.60 \\
        & GELU/SiLU & 7.95 & 17.17 \\     
        & Layer Norm & 0.98 & 0.71\\ \vspace{0.05in}
        & MatMul Truncations & 83.64 & 79.53 \\ \midrule
        \multirow{4}{*}{ARC-Easy} & Softmax & 4.41 & 0.80 \\
        & GELU/SiLU & 71.77 & 84.99 \\     
        & Layer Norm & 8.83 & 3.49\\ \vspace{0.05in}
        & MatMul Truncations & 15.00 & 10.72 \\
        \multirow{4}{*}{HellaSwag} & Softmax & 8.86 & 1.61 \\
        & GELU/SiLU & 68.43 & 84.29 \\     
        & Layer Norm & 8.42 & 3.46\\ \vspace{0.05in}
        & MatMul Truncations & 14.29 & 10.63 \\
        \multirow{4}{*}{MMLU} & Softmax & 17.84 & 3.40 \\
        & GELU/SiLU & 61.68 & 82.76 \\     
        & Layer Norm & 7.59 & 3.40\\ \vspace{0.05in}
        & MatMul Truncations & 12.89 & 10.44 \\
        \multirow{4}{*}{WinoGrande} & Softmax & 3.02 & 0.48 \\
        & GELU/SiLU & 72.81 & 85.26 \\     
        & Layer Norm & 8.96 & 3.50\\ 
        & MatMul Truncations & 15.21 & 10.76\\ \midrule
        \multirow{4}{*}{WikiText-2} & Softmax & 65.18 & 58.94 \\
        & GELU/SiLU & 26.14 & 35.18 \\     
        & Layer Norm & 3.22 & 1.45\\    
        & MatMul Truncations & 5.46 & 4.44 \\      \bottomrule
        \end{tabular}%
    }
    \centering
    \caption{\sl Percentage of truncation time taken up by the following nonlinearities: softmax, GELU/SiLU, Layer Norm. The remainder of the truncation time is taken up by in-field operations not directly related to one of the above approximations.}
    \label{tab:op_break}
\end{table}

\begin{figure}[ht]
    \centering
    \begin{minipage}{\columnwidth}
        \centering
        \includegraphics[width=\textwidth]{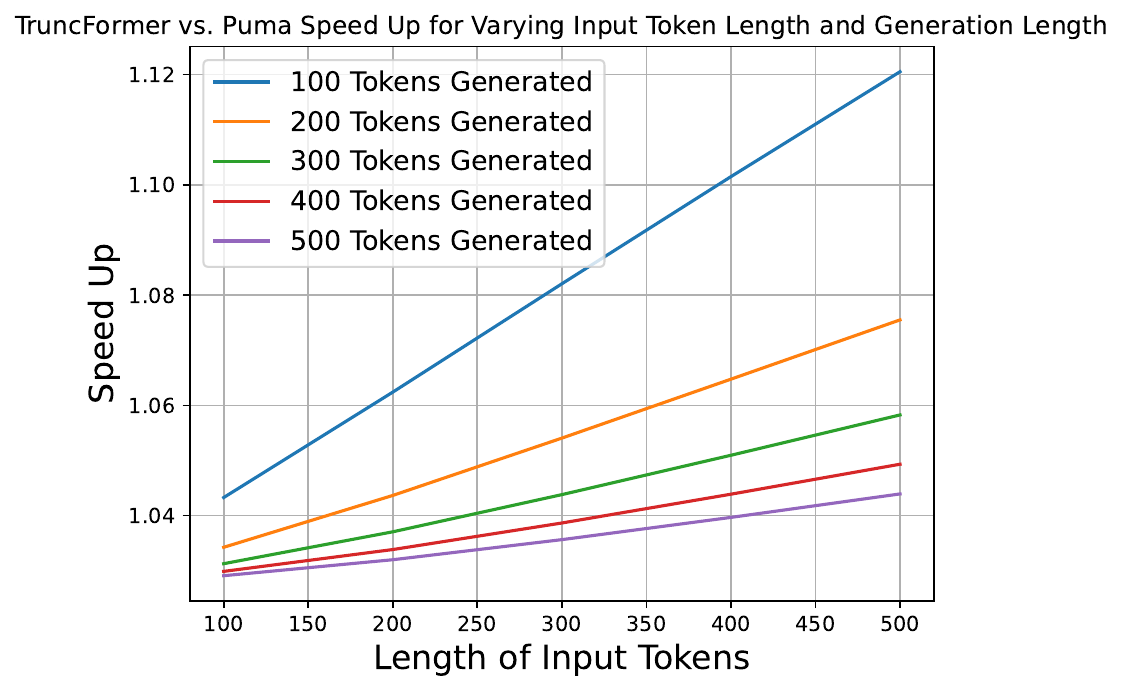}
        \caption{\small\sl We compare the impact of input length and token generation count on private inference latency.}
        \label{fig:input_plot}
    \end{minipage}
    
\end{figure}

We show empirical support of this theory in Figure~\ref{fig:input_plot}. In this plot we see that TruncFormer's speed up over Llama is dependent on both the length of the input tokens and the number of tokens generated. We note that there is a positive slope for all experiments when it comes to increasing the length of the input tokens. However, we see that the line with the largest slope is the one generating the smallest number of tokens. This suggests that the GELU/SiLU approximations do in fact play a strong role when it comes to comparing TruncFormer and Puma's latencies.

\paragraph{Impact of Larger Field Sizes on Latency}
In the previous experiments we have focused on a fixed field of size 64 bits. It becomes clear that with a larger field size, one can further reduce the number of truncations performed. However, a larger field also means larger circuits for performing the operations which leads to higher latency costs. Thus, we study the effects of a larger field size on truncation latency for Arc-Easy and MMLU. In particular, we compare a 64 bit field and a 128 bit field. Results are shown in Table~\ref{tab:field_size}.

We see that despite the larger field size, we do not see significant latency benefits. The reason for this likely lies in the aforementioned larger circuits. Our bench-marking time for a scalar truncation increases by around twofold when going from 64 bit circuits to 128 bit circuits (0.099ms to 0.175ms). Therefore, despite the number of truncations significantly decreasing, we spend a longer time on each truncation. This leads to minimal speedup despite a larger field size. 

\begin{table}[h]
    \resizebox{\columnwidth}{!}{%
        \begin{tabular}{cccc}\toprule
         Benchmark & Field Size (bits) & Accuracy (\%) & Speed-up ($\times$)\\ \midrule
         \multirow{2}{*}{ARC-Easy}& 128 & 73.02 & 1.04 \\ \vspace{0.05in} 
         &64 & 72.94 & \\ \midrule
          \multirow{2}{*}{MMLU}& 128 & 29.98 & 0.98 \\
         &64 & 29.92 & \\     \bottomrule
    \end{tabular}%
    }
    \centering
    \caption{\sl Performance for Llama-7B-T on ARC-Easy and MMLU given different field sizes. Speed ups are reported by taking the truncation time of the 64 bit model and dividing that by the time for the 128 bit model.}
    \label{tab:field_size}
\end{table}

\paragraph{Impact of Input Bit-Size on Accuracy}
We made the decision when performing our evaluations to set the input size to 13-bits. This choice was made because we found empirical results showing 13-bits as the lowest representation size to still provide comparable accuracy to the plaintext Llama-7B implementation. These results are shown across the Hellaswag and Wikitext-2-raw benchmarks in Table~\ref{tab:input_bench}. We see from the table that once we go below 13-bits the accuracy/perplexity has a drastic fall-off. One possible reason for this falloff has to do with our fixed point encoding described above. We multiply the floating point numbers by $2^n$ where n is the number of bits of precision we want. In the case of 13 bits, we have $2^{13} = 8192$ which is almost able to capture 4 decimal points of precision. Models like Llama-7B are commonly evaluated in floating point 16 which is also precise to 4 decimal points. However, we see that at 12 bits and below we start losing much of that precision which leads to a drastic drop in performance.

\begin{table}[ht]
    \resizebox{\columnwidth}{!}{%
        \begin{tabular}{ccccccc}\toprule
      Input Size (Bits) & Hellaswag (\%) & Wikitext-2-raw (perp.) \\ \midrule
      16 & 76.29 & 5.04\\
      15 & 76.30 & 5.04\\
    14 & 76.26 & 5.04\\
    \textbf{13} & \textbf{76.36} & \textbf{5.04}\\ \midrule
      12 & 59.48 & 8.68\\
      11 & 60.68 & 39.71\\
      10 & 34.01 & 27,122\\ \bottomrule
    \end{tabular}%
    }

    \centering
    \caption{\sl Performance for Llama-7B-T on Hellaswag and Wikitext-2-raw given a certain fixed point input size. The original Llama-7B has an accuracy of 76.20\% on Hellaswag and a perplexity of 5.04 on Wikitext-2-raw.}
    \label{tab:input_bench}
\end{table}

\paragraph{Impact of Iteration Count on Accuracy of Approximations}
Complex operations such as the softmax, SiLU, and Layer Norm can be written as a combination of simpler operations such as the exponential, reciprocal, and inverse square-root. However, all three of these "simpler" operations still require PI friendly approximations. Our approximations for these operations are based on the limit approximation or Newton-Raphson approximation methods. Therefore, the number of iterations plays a key role in the accuracy of these approximations. We study the impact of iteration count on these approximations by evaluating Llama-7B on the Arc-Easy test with various iteration counts. Each entry in Table~\ref{tab:approx_iter} states the perplexity on Wikitext-2-raw where the iteration count for only that specific operation has been changed. This helps us isolate the impact of iterations for different operations. We notice that the exponentiation and inverse square-root algorithms have perplexity degradations after a few reductions in iterations. However, the reciprocal operation does not seem to be affected to the same extent. 

\begin{figure}[!ht]
    \centering
    \begin{subfigure}{0.95\columnwidth}
        \centering
        \includegraphics[width=0.95\columnwidth]{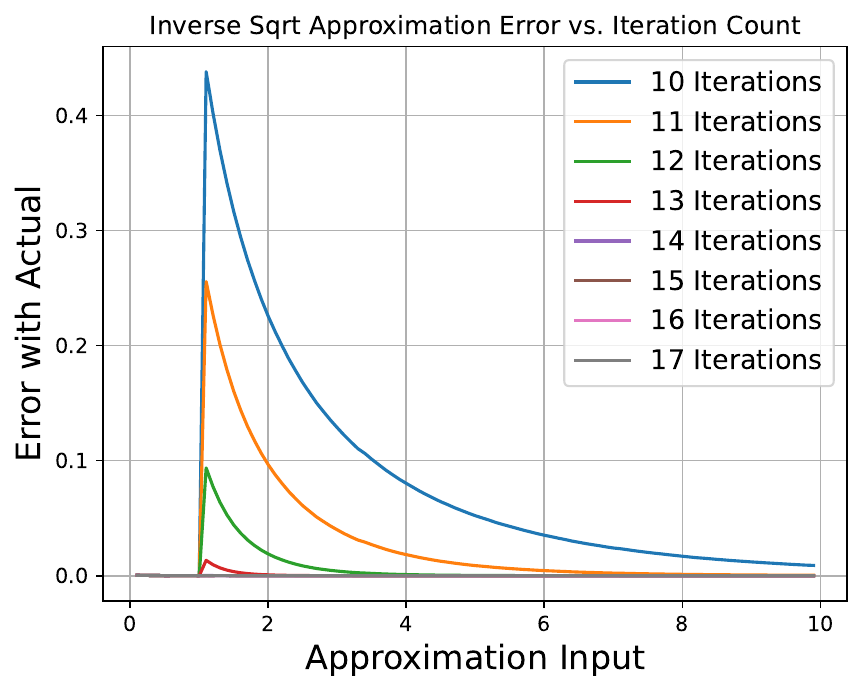} 
        \caption{Approximations for the inverse square-root with various iteration counts.}
        \label{fig:figure1}
    \end{subfigure}
    \begin{subfigure}{0.95\columnwidth}
        \centering
        \includegraphics[width=0.95\columnwidth]{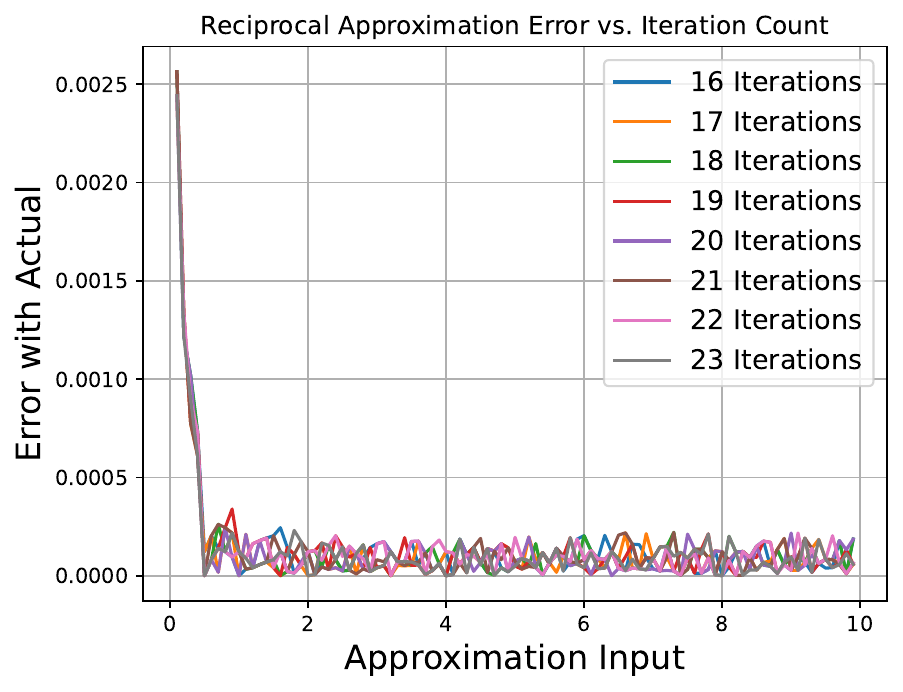} 
        \caption{Approximations for the reciprocal with various iteration counts.}
        \label{fig:figure2}
    \end{subfigure}
    \caption{We approximate the inverse-sqrt/reciprocal of the range of numbers (0, 10] with intervals of 0.1. The plots showcase the error between these approximations and the true function as defined by PyTorch. We vary the number of Newton-Raphson iterations for different plotted lines.}
    \label{fig:approx_plots}
\end{figure}

To further explore this, we perform micro-benchmarks on the accuracy of the approximations without considering downstream tasks. We approximate the inverse square-root/reciprocal of the range of numbers (0, 10] with intervals of 0.1. The plots showcase the magnitude of the error between these approximations and the true function as defined by PyTorch. We vary the number of Newton-Raphson iterations for different plotted lines. These plots are shown in Figure~\ref{fig:approx_plots}. In the inverse square-root plot we vary the number of iterations from 10 - 17. We see that the error is low for all of the iteration counts from the range of (0, 1), but then it spikes up to over 0.4 for the 10 iterations approximation. We see that as the iteration count increases, our approximation error decreases until it becomes near 0 at 14 iterations. The general trend holds for larger input values. The reciprocal however offers a different perspective. We see a high error at the beginning of the input range, but immediately after we have all of the approximations trend towards 0 error. This seems to hold for iteration counts both above and below 20 iterations. This suggests that there is much more flexibility in the number of iterations for the reciprocal than there is for the inverse square-root. This supports the results found in Table~\ref{tab:approx_iter} where a change in the reciprocal's iteration count has a much lower effect on the downstream perplexity than the inverse square-root. Future work in this area could lead to better initial estimates that can further lower the number of iterations necessary for accurate approximations.


\begin{table}[ht]
    \resizebox{\columnwidth}{!}{%
        \begin{tabular}{cccccc}\toprule
         & \multicolumn{5}{c}{Change in Iteration Count} \\ \midrule
        Operation & -4& -3& -2 & -1 & 0 \\ \midrule
        Exponentiation &5.35&5.29& 5.03& 5.24 &5.04\\
        Reciprocal&5.03& 5.03&5.03 & 5.03 &5.04 \\
        Inverse Square-Root& 6.70&5.25 &5.05 &5.04 &5.04\\ \bottomrule
    \end{tabular}%
    }
    \centering
    \caption{\sl Llama-7B performance on Wikitext-2-raw with varying iteration counts for each operation's approximation. Iterations are changed in isolation; for any entry in the table, the other two operations will have their change in iteration count be 0. The default iteration counts are 8 for exponentiation, 20 for reciprocal, and 12 for inverse square-root.}
    \label{tab:approx_iter}
\end{table}

\paragraph{Continued Impracticality of PI}
Throughout this work we have focused on relative latencies by comparing TruncFormer to Puma. However, you might be wondering how these results translate into actual time. Unfortunately, the runtime for PI is still in the hours for a single inference. Therefore, we note that PI is still impractical, even with TruncFormer. We propose a new research direction centered around optimizing truncations rather than constantly creating new approximations for other nonlinear functions. We believe this direction is promising because it lets researchers focus on the one operation that contributes to at least 90\% of the total private inference runtime.

\section{Conclusions}

In this paper we propose a framework for taking a machine learning model and statically determining the optimal location for truncations to improve the latency of private inference. Our evaluations show that our framework maintains plaintext accuracy while also significantly reducing the latency of private inference. In addition, we show that optimizing the truncation operation will lead to more significant latency gains. 

One limitation in our work is the focus on a specific cryptographic protocol (Delphi). We chose Delphi because it provided the easiest protocol for separating truncations and multiplications for latency calculations. We believe that this work should be applicable to other protocols due to truncations remaining an expensive operation in all popular protocols. However, we leave confirmation of this hypothesis for future work.


\bibliography{example_paper}
\bibliographystyle{mlsys2025}



\end{document}